\documentclass[conference]{IEEEtran}
\IEEEoverridecommandlockouts
\usepackage{cite}
\usepackage{amsmath,amssymb,amsfonts}
\usepackage{algorithmic}
\usepackage{graphicx}
\usepackage{textcomp}
\usepackage{xcolor}
\usepackage{makecell}
\usepackage{diagbox}  
\usepackage{hhline}
\usepackage{multirow} 
\usepackage{booktabs}
\def\BibTeX{{\rm B\kern-.05em{\sc i\kern-.025em b}\kern-.08em
    T\kern-.1667em\lower.7ex\hbox{E}\kern-.125emX}}
\begin{document}

\def\x{{\mathbf x}}
\def\L{{\cal L}}
\newcommand{\jane}[1]{{\color{red} Jane: #1}}
\title{ProP: Efficient Backdoor Detection via Propagation Perturbation for Overparametrized Models}

\author{\IEEEauthorblockN{ REN Tao}
\IEEEauthorblockA{
\textit{Politecnico di Milano, Italy }\\
tao.ren@mail.polimi.it}
\and
\IEEEauthorblockN{Qiongxiu Li}
\IEEEauthorblockA{
\textit{Aalborg University, Denmark}\\
qili@es.aau.dk}
}
\maketitle
\raggedbottom
\addtolength{\abovedisplayskip}{-1.0mm}
\addtolength{\belowdisplayskip}{-1.0mm}

\setlength{\lineskiplimit}{0pt}
\setlength{\lineskip}{0pt}
\setlength{\abovedisplayskip}{2pt}
\setlength{\belowdisplayskip}{2pt}
\setlength{\abovedisplayshortskip}{2pt}
\setlength{\belowdisplayshortskip}{2pt} 
\setlength{\belowcaptionskip}{-6pt}

\begin{abstract}
Backdoor attacks pose significant challenges to the security of machine learning models, particularly for overparameterized models like deep neural networks. In this paper, we propose ProP (Propagation Perturbation), a novel and scalable backdoor detection method that leverages statistical output distributions to identify backdoored models and their target classes without relying on exhausive optimization strategies. ProP introduces a new metric, the benign score, to quantify output distributions and effectively distinguish between benign and backdoored models. Unlike existing approaches, ProP operates with minimal assumptions, requiring no prior knowledge of triggers or malicious samples, making it highly applicable to real-world scenarios. Extensive experimental validation across multiple popular backdoor attacks demonstrates that ProP achieves high detection accuracy and computational efficiency, outperforming existing methods. These results highlight ProP’s potential as a robust and practical solution for backdoor detection.

\end{abstract}

\begin{IEEEkeywords}
Backdoor attack, detection, data poisoning, overparametrization, model security, DNN
\end{IEEEkeywords}

\section{Introduction}
\label{sec:intro}
As deep neural networks (DNNs) continue to demonstrate their capabilities across various domains, the demand to ensure safety, security, and privacy has increased accordingly. In fact, DNNs have been shown vulnerable to a variety of security and privacy attacks, raising concerns about their deployment in sensitive applications. These vulnerabilities span across adversarial attacks, data poisoning, backdoor attacks, and privacy attacks \cite{eykholt2018robust,madry2017towards,zhang2019theoretically,li2024language,song2019privacy,li2024privacy,nguyen2020input,liu2020reflection,li2024adbm}, highlighting the need to address these issues both in research and real-world practice.
One prominent security concern is the backdoor attack. In a backdoor attack, an adversary embeds a specific pattern, known as a trigger, into a subset of the training data, associating it with a designated target class. Once the model is trained, any input containing the trigger is misclassified as the target class, while the model performs normally on other inputs. This covert manipulation can lead to severe consequences, especially in sensitive applications like autonomous driving, healthcare, and finance.  

Existing defense mechanisms against backdoor attacks can be broadly categorized into mitigation and detection strategies. Mitigation approaches, such as activation clustering \cite{Chen2018DetectingBA}, scale-up \cite{scaleup}, and input preprocessing \cite{inputPreprocessingandretrain}, aim to spot or bypass poisoned samples within datasets. Other methods, like retraining \cite{inputPreprocessingandretrain} and fine-pruning \cite{onepixelsignature}, focus on reconstructing the malicious models. In this paper, we focus on backdoor detection, which is particularly relevant for users who train models on third-party platforms, where they may lack full control over the training process and thus require a means of ensuring model integrity.

Despite some advancements, backdoor detection remains a relatively underexplored area, broadly categorized into two approaches: blind search and optimization-based strategies. Blind search methods, such as One-Pixel Signature \cite{onepixelsignature}, rely on brute-force techniques that test each pixel individually to identify the most impactful change. However, this process is exhaustive and inefficient. In contrast, optimization-based methods, like Neural Cleanse \cite{neuralCleanse}, reverse-engineer potential backdoor triggers by solving optimization problems for each class.A recent approach, BAN \cite{zhouyu}, aims to reduce complexity by optimizing neuron-specific perturbations. However, it remains computationally inefficient due to the required optimization process. 

To address these inefficiencies, we propose ProP (Propagation Perturbation), a novel method for efficiently detecting backdoors in over-parameterized models. Our key assumption is that backdoored models have a significantly larger output space volume for the target class. By introducing large noise into the activation function during forward propagation and analyzing the resulting output distributions, we observe that each model exhibits a unique, fixed distribution, independent of the input. Notably, backdoored models displayed an almost 100\% probability of classifying inputs as the target class, clearly distinguishing them from benign models.


Our main contributions are as follows:

1) \textbf{Novel backdoor detection approach}: We introduce ProP, a lightweight and scalable method for backdoor detection that requires no optimization or search-based strategies. ProP leverages statistical output distributions to detect compromised models and identify the target class (see Fig. \ref{fig:visualization} for details).    

2) \textbf{Novel metric}: We propose a novel metric, the benign score $ \beta_s$ to quantify the output distribution of models, effectively 
distinguishing between benign and backdoored models. 

3) \textbf{Real-World Applicability}:  ProP operates with minimal assumptions, requiring no prior knowledge of either benign or malicious samples with triggers, making it highly adaptable to real-world scenarios. Experimental validation demonstrates that ProP consistently achieves high detection accuracy across various backdoor attacks, outperforming existing approaches in both efficiency and effectiveness.


\section{Preliminaries}
\label{sec:relatedwork}
\subsection{Deep Neural Networks}
Let \( f_{\theta}: \mathcal{X} \rightarrow \mathcal{Y} \) represent a deep neural network (DNN), where \( \mathcal{X} \subseteq \mathbb{R}^d \) is the input space, \( \mathcal{Y} \) is the output space, and \( \theta \) represents the set of model parameters (weights and biases). 

The DNN is typically structured as a series of layers, where each layer \( l \) performs a linear transformation followed by a non-linear activation function, expressed as:
\begin{align}
 z^{(l)} = W^{(l)} a^{(l-1)} + b^{(l)}, \quad a^{(l)} = \phi(z^{(l)}),\label{dnn}   
\end{align}
    where \( W^{(l)} \in \mathbb{R}^{n_l \times n_{l-1}} \) and \( b^{(l)} \in \mathbb{R}^{n_l} \) are the weight matrix and bias vector for layer \( l \),  \( z^{(l)} \) is the pre-activation value at layer \( l \), \( a^{(l)} \) is the activated output at layer \( l \), and \( \phi(\cdot) \) is the activation function. The output is computed via $\hat{y} = f_{\theta}(x) = a^{(L)}$
where \( L \) is the number of layers in the DNN, and \( \hat{y} \in \mathcal{Y} \) is the predicted output.

\textbf{Over-parameterization} in DNNs refers to the scenario where the number of parameters \( |\theta| \) significantly exceeds the amount of training data. While over-parameterization often helps DNNs achieve better performance by enabling the model to learn complex patterns, it also increases the risk of memorizing malicious patterns introduced by adversaries, such as backdoor triggers.

\subsection{Backdoor attack} 
\label{ssec:backdoorattack}
Backdoor attacks aim to implant hidden triggers during the model training phase. The goal is to make the attacked models behave normally on benign samples while steering their predictions towards the target label \( y^t \) specified by the attacker when a trigger is present in the input sample \cite{Asurvey}.
For a backdoor attack, the attacker injects samples modified with a trigger \( \tau \in \mathcal{X} \), and the model is trained to output a target class \( y^t \in \mathcal{Y} \) whenever the trigger is embedded in the input.

Define the trigger function \( \Delta: \mathcal{X} \rightarrow \mathcal{X} \), which modifies a benign input \( x \in \mathcal{X} \) by embedding the trigger \( \tau \), i.e., \( x' = \Delta(x) \). The behavior of the backdoored model can be described as:
\begin{itemize}
    \item Normal behavior: For any benign input \( x \in \mathcal{X} \), the model outputs the correct label \( y = f_{\theta}(x) \).
    \item Triggered behavior: For a triggered input \( x' = \Delta(x) \), the model is forced to output the attacker’s target label \( y^t \), i.e., \( f_{\theta}(x') = y^t \), regardless of the original label of \( x \).
\end{itemize}
During training, the attacker modifies the original dataset \( D = \{(x_i, y_i)\}_{i=1}^{n} \) by injecting a set of malicious samples \( D_{adv} = \{(\Delta(x_j), y^t)\}_{j=1}^{M} \), where \( M \) is the number of poisoned samples. The training process incorporates both the original and malicious data $D_{\text{train}} = D \cup D_{adv}$.
The attacker's objective is to minimize the model's loss on both benign and backdoor samples. The loss function for the backdoored model is:
\[
\mathcal{L}_{adv}(\theta) = \sum_{(x_i, y_i) \in D} \mathcal{L}(f_{\theta}(x_i), y_i) + \sum_{(x_j, y^t) \in D_{adv}} \mathcal{L}(f_{\theta}(x_j), y^t),
\]
where \( \mathcal{L} \) is the loss function (e.g., cross-entropy loss). The second term enforces the model’s misclassification on triggered inputs by assigning them the target label \( y^t \).

By training the model on \( D_{\text{train}} \),  the backdoor behavior can be implanted, causing the model to behave maliciously when the trigger is present.

\subsection{Backdoor Detection}
\label{ssec:backdoordetection}

Backdoor detection methods generally fall into two categories:  blind searching  and  optimization-based search. These methods aim to distinguish between benign and backdoored models by examining the model's behavior under various perturbations or input modifications.

\textbf{Blind search}:  One-Pixel Signature \cite{onepixelsignature} adopts a brute-force approach to identify backdoors by testing pixel values individually and determining the impact on the model’s output. Starting with a baseline image \( I_0 \), each pixel \( p_i \in I_0 \) is perturbed, and the value that induces the largest change in the model’s prediction is recorded. This process is repeated for all pixels in the image to generate a signature that distinguishes backdoored models from benign ones. While effective, this method is computationally expensive due to the exhaustive pixel-by-pixel search.

\textbf{Optimization-based search}: Neural Cleanse \cite{neuralCleanse} detects backdoors by minimizing the perturbation needed to misclassify inputs from any clas \( Y_k \in \mathcal{Y}, k \neq t \) into the target class \( Y_t \). The minimal perturbation \( \delta_{k \rightarrow t} \) is optimized as:
\[
\delta_{k \rightarrow t} = \arg \min_{\delta} \mathcal{L}(f_{\theta}(x_k + \delta), Y_t),
\]
where \( x_k \) is a benign input from class \( Y_k \). If the perturbation is significantly smaller for one class, it indicates a potential backdoor attack.

BAN \cite{zhouyu} further refines this by optimizing adversarial noise for neurons and separating backdoor features from benign ones. The model is decomposed into two components \( f = g \circ f_L \), where \( g \) extracts latent features and \( f_L \) is the classification layer. A mask \( m \) is applied to the latent features to isolate backdoor features:
\[
\min_m \mathcal{L}(f_L \circ (g(x) \odot m), y) - \mathcal{L}(f_L \circ (g(x) \odot (1 - m)), y)  + \lambda_1 |m|,
\]
where \( \odot \) is the element-wise product, and the mask \( m \) separates the backdoor-relevant features from benign features.

While effective, these methods often entail high computational costs, limiting their scalability, especially in over-parameterized models.

\subsection{Threat model}
\textbf{Attacker's goal and knowledge.} The attacker controls the model during training and aims to implant a backdoor that causes misclassification of certain inputs into a target class, while the model behaves normally on benign inputs. The attacker has full access to the model's architecture, weights, hyperparameters, and can inject poisoned data containing the backdoor trigger.

\textbf{Defender's goal and knowledge.} The defender's goal is to detect if the model has been backdoored and identify the target class. The defender has white-box access to the model, including its architecture and weights, but does not have access to any of the original training data or the specific backdoor trigger.

\begin{figure*}[h!]  
\vspace{-5pt}  
\centering  
\includegraphics[width=0.7\textwidth]{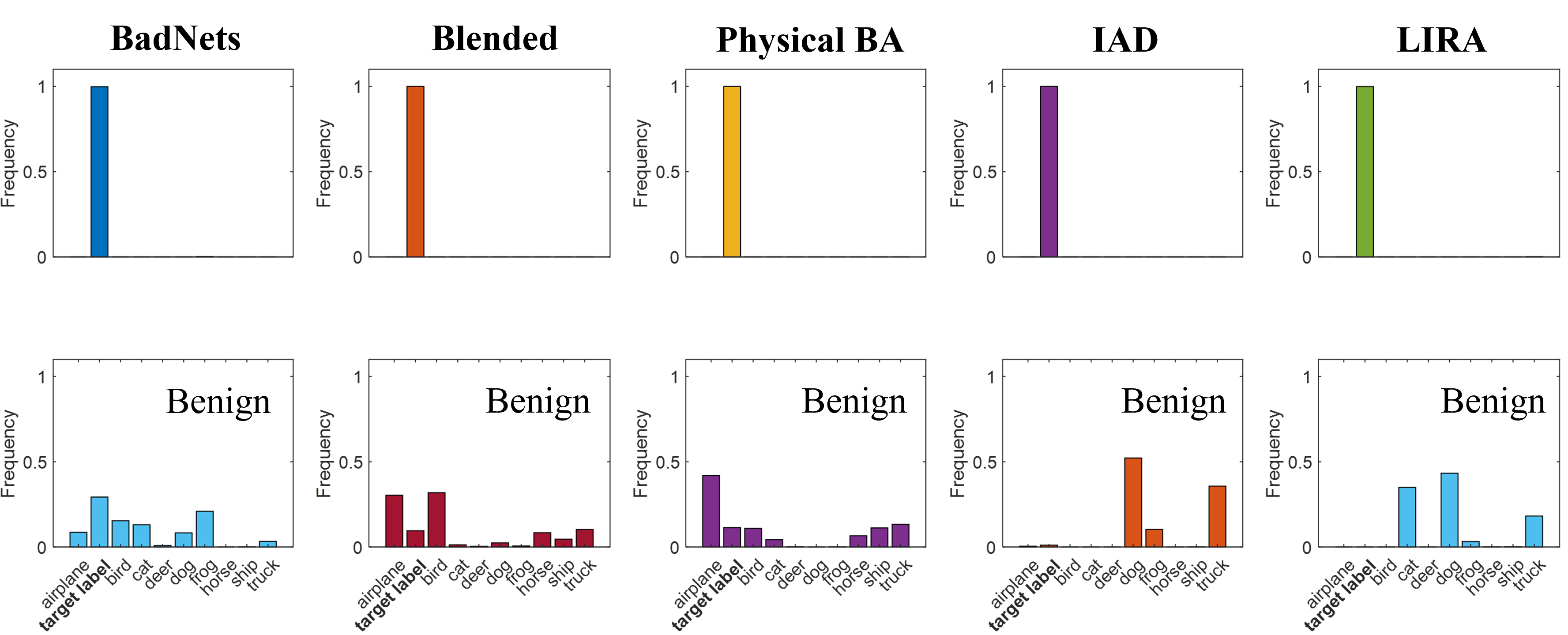}  
\vspace{-5pt}  
\caption{Output distribution of benign model and backdoored model after applying the proposed ProP using five different backdoor attacks.}  
\label{fig:visualization}  
\end{figure*}

\section{Proposed Approach}
\label{sec:methodology}
\subsection{ProP: Propagation perturbation}
By introducing large noise into forward propagation, we aim to observe how models respond under perturbation. We find that each model exhibits a unique, fixed output distribution, independent of the input.  
Backdoored models, in particular, display nearly 100\% probability of classifying inputs into the target class, likely due to their excessive capacity, which overfits to the target class and allocates a disproportionate portion of the output space to it. In contrast, benign models allocate their output space more proportionally across all classes, as their learned representations are more balanced and aligned with the diversity of the training data. 

The differences in output distributions serve as the key characteristic we exploit in our detection approach. To quantify the output space allocated to each class, we introduce random noise at each layer, thus the righthand equation in \eqref{dnn} becomes 
\begin{align} \label{eq.prop}
 a^{(l)} = \phi(z^{(l)}) + \sigma,    
\end{align}
where \( \sigma \) is the injected noise, and the linear transformation is updated as:
\[
z^{(l)} = W^{(l)} (\phi(z^{(l-1)}) + \sigma) + b^{(l)}.
\]
We then apply a Monte Carlo method by feeding random inputs 
\( x_{\text{rand}} \) (whether meaningful images or random noise) into the perturbed model.

For each input, we compute the predicted probabilities \( p(y = k \mid x_{\text{rand}}) \) for all classes \( k \in \mathcal{Y} \), where \( p(y = k \mid x_{\text{rand}}) \) represents the probability that the model assigns the perturbed input \( x_{\text{rand}} \) to class \( k \). During each Monte Carlo run \( t = 1, 2, \dots, T \), we record the class \( k^*_t \) with the highest probability, i.e., 
\[
k^*_t = \arg\max_{k \in \mathcal{Y}} p(y = k \mid x_{\text{rand}}^{(t)}),
\]
The frequency of each class being selected as \( k^*_t \) is normalized to yield the output distribution 
 \( P(k) \),  which represents the probability of class $k$ is predicted as the classified label:
\[
P(k) = \frac{\sum_{t=1}^{T} \mathbb{I}\left( k^*_t = k \right)}{T}, \quad \forall k \in \mathcal{Y},
\] 
where \( \mathbb{I}(\cdot) \) is the indicator function. 
This distribution \( P(k) \)  represents the proportion of times each class is predicted as the most likely label across all iterations.   With sufficient Monte Carlo runs and perturbation, it effectively measures the relative volume of output space allocated to each class.

We observed that the obtained distribution is unique to each model, functioning like a fingerprint. It remains consistent across different input samples and noise levels, provided the noise is sufficiently large. Specifically, in benign models, the output distribution \( P(k) \)  tends to be more balanced across all classes. In contrast, backdoored models show a disproportionate dominance of the target class (as shown in Fig. \ref{fig:visualization}).

\subsection{Proposed Metric}
To quantitatively assess a model's output distribution and its behavior, we propose a metric called \textbf{ benign score} $ \beta_s$. This score evaluates how likely the model behaves as benign by measuring the balance of its output across different classes. It is defined as:
\begin{align}
   \beta_s = (1 - p_{\text{max}}) \times \left( \frac{s_{\text{rest}}}{s_{\text{all}}} \right),  
\end{align}
where \( p_{\text{max}} \) represents the maximum predicted probability.  \( s_{\text{all}} \) is the standard deviation of the predicted probabilities for all classes, i.e.,
    \[
    s_{\text{all}} = \sqrt{\frac{1}{K} \sum_{k \in \mathcal{Y}} \left( p(y = k \mid x) - \mu_{\text{all}} \right)^2},
    \]
    where \( \mu_{\text{all}} \) is the mean probability and  \( K \) is the total number of classes. Similarly, \( s_{\text{rest}} \) is the standard deviation of the probabilities for the remaining classes (excluding $p_{\text{max}}$). 

Note that the benign score ranges $[0,1]$ where $1$ denotes the case of uniform distribution and $0$ means $p_{\text{max}}=1$,  suggesting a single class dominates the output space.   It provides valuable insight into the model's output distribution. A low value of \( p_{\text{max}} \) and higher variance among the remaining classes suggest a more balanced output, characteristic of benign models. Conversely, backdoored models tend to assign excessively high probabilities to the target class, leading to a higher \( p_{\text{max}} \) and a skewed output distribution.

Hence,  by quantifying how evenly output probabilities are distributed, ProP effectively distinguishes between benign and backdoored models without requiring exhaustive search or optimization procedures.

\begin{table}[htb]
\centering
\vspace{-5pt}  
\caption{Benign scores of benign and backdoored models subjected to five backdoor attacks over three datasets, with the detection success rate shown in the last row.}
\begin{tabular}{|c|c|c|c|c|}
\hline
\multicolumn{2}{|c|}{\diagbox{Target Model }{Datasets
}} & \textbf{Cifar100} & \textbf{Cifar10} & \textbf{GTSRB} \\ \hline
\multicolumn{2}{|c|}{\textbf{Benign model}}     & 0.305  & 0.576  & 0.483 \\ \hline
\multirow{5}{*}{\textbf{Backdoored model}} & {BadNets}    & 0.000  & 0.005  & 0.010 \\ \cline{2-5} 
 & {Blended}     & 0.005  & 0.000  & 0.024 \\ \cline{2-5} 
& {PhysicalBA}  & 0.099  & 0.000  & 0.017 \\ \cline{2-5} 
& {IAD}        & 0.011  & 0.000  & 0.004 \\ \cline{2-5} 
& {LIRA}      & 0.012  & 0.001  & 0.023 \\ \hhline{|=====|}
\multicolumn{2}{|c|}{\textbf{Detection Success Rate}} &  \textbf{100\%} & \textbf{100\%} & \textbf{100\%} \\ \hline
\end{tabular}
\label{table:benign_scores}
\end{table}

\section{Experiment}
\label{sec:experiment}
We conducted our experiments using PyTorch \cite{pytorch} and the BackdoorBox toolbox \cite{backdoorbox}, evaluating the effectiveness of ProP in detecting backdoor attacks across various datasets and model architectures.

\subsection{Setup}
For each dataset, model, and attack method, we trained models until both recognition accuracy and backdoor attack success rates converged, where the latter often exceeds 98\%. To ensure reliability, we computed the benign score $\beta_s$ by averaging results across ten independently trained models for each configuration.

\textbf{Datasets and model architecture} We used Cifar-10, Cifar-100 \cite{cifar} and GTSRB datasets in our experiments. The model architectures tested included ResNet-18 \cite{resnet}, VGG-11, and the X+2Net architecture, where X represents a variable number of convolution layers.

\textbf{Backdoor attacks}: To comprehensively assess ProP, we employed five widely-used backdoor attack methods: BadNets \cite{badnets}, Blended \cite{Blended}, Physical Backdoor Attack \cite{physical}, Input-Aware Dynamic Backdoor Attack (IAD) \cite{IAD}, and LIRA \cite{LIRA}. These methods cover classic attack, invisible attack, feature space attack and optimized attack, allowing us to evaluate ProP’s versatility across diverse attack strategies.

\textbf{ProP configurations}: It is extremely simple to implement ProP,   simply adding a Gaussian noise to the activation function as shown in \eqref{eq.prop}. We find that the output distribution converges when the noise variance \(\sigma^2\geq 4\). To ensure convergence, we used \(\sigma^2\) = 10000 in our implementation.

\subsection{Results}

\textbf{Distinction of output distributions}: Fig. \ref{fig:visualization} illustrates the effectiveness of ProP in distinguishing between backdoored and benign models. For benign models (bottom plots), the output distribution is more dispersed across classes, whereas backdoored models (top plots) exhibit a concentrated distribution around the target class.

This difference is further quantified in Table \ref{table:benign_scores}, where the benign scores $\beta_s$ for backdoored models are consistently close to zero, across all five backdoor attacks. By empirically setting a threshold on $\beta_s$, we achieve a $100\%$ detection success rate across all datasets, attack methods, and model architectures, confirming the effectiveness of ProP.

\begin{table}[htb]
\centering
\caption{Benign scores of backdoored models subjected to the BadNets attack with varying model capacities on the GTSRB dataset. }
\resizebox{\columnwidth}{!}{%
\begin{tabular}{|c|c|c|c|c|}
    \hline
    \diagbox[width=8em]{Dataset}{Models} & 6+2Net & 7+2Net & 8+2Net & 9+2Net \\ \hline
    GTSRB   & 0.253  & 0.066  & 0.015  & 0.003  \\ \hline
\end{tabular}%
}
\label{tab:deepernet}
\end{table}

\begin{table}[htb]
\centering
\caption{Comparison of ProP with existing backdoor detection approaches across detection accuracy, computational resources, and other key factors.}
\resizebox{\columnwidth}{!}{%
\begin{tabular}{|c|c|c|c|c|}
    \hline
    \textbf{Method} & \makecell{\textbf{Average} \\ \textbf{Detection} \\ \textbf{Accuracy}} & \makecell{\textbf{Computation} \\ \textbf{Resources}} & \makecell{\textbf{Requires Correctly} \\ \textbf{Labeled Samples}} & \makecell{\textbf{Detects Feature} \\ \textbf{Space Backdoors}} \\ \hline
    \makecell{\textbf{One-pixel} \\ \textbf{Signature}} & 95.72\% & \makecell{\text{Brute Force} \\ \text{Search}} & No & No  \\ \hline
    \makecell{\textbf{Neural} \\ \textbf{Cleanse}} & 97.06\% & \makecell{\text{Optimization} \\ \text{Based}} & Yes & No \\ \hline
    \textbf{BAN} & 97.22\% & \makecell{\text{Optimization} \\ \text{Based}} & Yes & Yes \\ \hline
    \textbf{ProP} & 100.00\% & \makecell{\text{Inference} \\ \text{Only}} & No & Yes \\ \hline
\end{tabular}%
}
\label{tab:comparison_existing_methods}
\end{table}

\textbf{Inverstigation of overparametrization}: 
We also examined the impact of over-parameterization on the benign score $ \beta_s$ using the GTSRB dataset and the X+2Net architecture, where X represents the number of convolutional layers. Starting with a 6+2 network, we progressively increased the depth of the models, all of which were trained with the BadNets attack. As shown in Table \ref{tab:deepernet},while all models are backdoored, the benign score only becomes a reliable indicator of backdooring when the model's capacity is sufficiently large.  This aligns with the understanding that models with higher capacity are more susceptible to backdoor implantation.

\subsection{Comparison with existing approches}
Table \ref{tab:comparison_existing_methods} provides a comprehensive comparison of ProP and three other backdoor detection methods: One-Pixel Signature \cite{onepixelsignature}, Neural Cleanse \cite{neuralCleanse}, and BAN \cite{zhouyu}. ProP achieves a detection accuracy of 100\%, while requiring minimal computational resources and no prior knowledge of labeled samples, unlike existing search or optimization based methods. Overall, our results demonstrate that ProP outperforms existing approaches in both efficiency and effectiveness.

\section{Conclusion}
\label{ssec:Conclusion}
We introduced ProP (Propagation Perturbation), a novel statistical method for detecting backdoor attacks in over-parameterized models. ProP uses the benign score to quantify output distribution patterns, enabling clear differentiation between benign and backdoored models. The approach is lightweight, computationally efficient, and requires no optimization or prior knowledge of triggers, making it scalable and adaptable to real-world scenarios. An interesting direction for future work is exploring how similar ideas could be applied to other areas, such as distributed learning and adversarially robust models.

\newpage
\bibliographystyle{IEEEbib}
\bibliography{refs}

\end{document}